\documentclass[twoside,leqno,twocolumn]{article}
\usepackage{ltexpprt}
\DeclareMathAlphabet\mathbfcal{OMS}{cmsy}{b}{n}
\usepackage[skip=0pt]{caption}
\usepackage{amsmath}
\usepackage{xcolor}
\usepackage{algorithm}
\usepackage{algorithmic}
\usepackage{multirow}
\usepackage{enumitem}
\usepackage{longtable}
\usepackage{amssymb}
\usepackage{graphicx}
\usepackage{float}
\usepackage{balance}
\usepackage{changepage}
\usepackage{tabularx}
\usepackage{amsmath}
\usepackage{amsfonts}
\usepackage{amssymb}
\usepackage{color}
\usepackage{xcolor,colortbl}

\begin{document}

\title{\Large Discovering Hidden Topical Hubs and \\ Authorities in Online Social Networks \thanks{This work is supported by the National Research Foundation under its International Research Centre@Singapore Funding Initiative and administered by the IDM Programme Office, and National Research Foundation (NRF). This is also supported by the ERC Grant (339233) ALEXANDRIA.}}
\author{Roy Ka-Wei Lee\thanks{Living Analytics Research Centre, Singapore Management University. \{roylee,eplim\}@smu.edu.sg} \\
\and
Tuan-Anh Hoang\thanks{L3S Research Center, Leibniz University of Hanover, hoang@l3s.de}
\and 
Ee-Peng Lim\footnotemark[2]}
\date{}

\maketitle

\fancyfoot[R]{\footnotesize{\textbf{Copyright \textcopyright\ 2018 by SIAM\\
Unauthorized reproduction of this article is prohibited}}}

\begin{abstract} \small\baselineskip=9pt Finding influential users in online social networks is an important problem with many possible useful applications. HITS and other link analysis methods, in particular, have been often used to identify hub and authority users in web graphs and online social networks. These works, however, have not considered topical aspect of links in their analysis. A straightforward approach to overcome this limitation is to first apply topic models to learn the user topics before applying the HITS algorithm. In this paper, we instead propose a novel topic model known as Hub and Authority Topic (HAT) model to combines the two process so as to jointly learn the hub, authority and topical interests. We evaluate HAT against several existing state-of-the-art methods in two aspects: (i) modeling of topics, and (ii) link recommendation. We conduct experiments on two real-world datasets from Twitter and Instagram. Our experiment results show that HAT is comparable to state-of-the-art topic models in learning topics and it outperforms the state-of-the-art in link recommendation task.\end{abstract}

\section{Introduction}
\label{sec:introduction}
\textit{Motivation.} Online social networks (OSNs), such as Facebook, Twitter and Instagram, have grown monumentally over recent years. It was reported that as of August 2017, Facebook has over 2 billion monthly active users, while Instagram and Twitter have over 700 million and 300 million monthly active user accounts respectively \cite{statista2017}. The vast amount of content and social data gathered in these behemoth platforms have made them important resources for marketing campaigns such as gathering consumer opinions and promotion of new products. Identifying influential users in OSNs is critical to such marketing activities.

Many research works have proposed methods to identify influential users in online social networks. For example, there are studies \cite{gayo2013, romero2011, shahriari2014} that adapted the HITS algorithm, which was originally proposed to extract user information by analyzing link structure in the World Wide Web \cite{kleinberg1999}, to identify influential users in OSNs. Gayo-Avello \cite{gayo2013} applied HITS on Twitter users' following-follower relationships to identify and differentiate influential users from spammers; the work considers an \textit{authority} user to be someone who is followed by many \textit{hub} users, while a \textit{hub} user to be one who follows many \textit{authority} users. Romero et al. \cite{romero2011} proposed the influence-passivity (I-P) algorithm, which is closely related to HITS, to measure user influence and passivity from their retweet activities in Twitter. Shahriari and Jalili \cite{shahriari2014} further extended HITS to identify influential users in signed social networks. Nevertheless, these existing works do not consider the topic specificity of the user links or activities when applying HITS. 

Topic specificity is important when analyzing the hub and authority users as it provides the context to user hub or authority. Consider an example of two users, $u_{1}$ and $u_{2}$, sharing similar ego network structures. HITS will therefore assign the two users similar authority and hub scores. However if $u_{1}$ is a popular food content contributor who is followed by many food-loving users, while $u_{2}$ is a prominent politician followed by many users interested in politics. As such, $u_{1}$ and $u_{2}$ are authority users on food-related and political topics respectively.

The benefits of studying topic-specific hub and authority users are manifold. Firstly, it enables better user recommendation. For example, we can recommend a jazz-loving user to follow another user \textit{v}, who is an authority user in jazz music. Secondly, identifying topic-specific hub and authority users enhances the effectiveness of marketing campaign. For example, a food and beverage company can reach out to food-related topics authority users to promote their products. It can also find new food authorities by tracking the hub users interested in food topics.

\textit{Research Objectives and Contributions.} In this paper, we aim to model topic-specific hub and authority users in OSNs. A straightforward approach to overcome this limitation is to first apply topic models such as LDA \cite{blei2003} to learn the users' topics before applying HITS. However, this two-step approach is non-optimal. Thus, we propose Hub and Authority Topic (HAT) model to unify the two steps into one that  to jointly learns the hub, authority and topical interests of users simultaneously. 

In our research, we first develop a generative story for both the content and links in an online social network so as to define our proposed Hub and Authority Topic (HAT) model. We then work out the parameter learning steps.  To evaluate the HAT model, we collect users' link and post data from Instagram and Twitter. We perform two sets of experiments to evaluate HAT: (i) we use likelihood and perplexity to evaluate the model's ability in learning topics from user generated content, and (ii) we evaluate our model ability to recommend topical influential users through user link prediction.

This paper improves the state-of-the-art by making two main contributions. Firstly, to the best of our knowledge, HAT is the first model that jointly learns user topics, hub and authority in social networks. In contrast, many of the previous works either study hub and authority users that are  topic-oblivious or model the topics, hub and authority separately. Secondly, through experimentation with two real-world datasets, we demonstrate (a) HAT is comparable to state-of-the-art topic models in learning topics from user generated content, and (b) HAT outperforms other models in user link recommendation.

The rest of this paper is organized as follows: We first discuss the related works in Section \ref{sec:related}. We then present the Hub and Authority Topic (HAT) model in Section \ref{(sec:model)}. Section \ref{(sec:experiment} presents the real-world data and experimental evaluations. The empirical study on the real-world data using our model will also be discussed. Finally, we conclude the paper and discuss the future works in Section \ref{(sec:conclusion}.

\section{Related Works}
\label{sec:related}
We broadly classify the existing research on influential social media users into two categories: topic-oblivious and topic-specific. For topic-oblivious works, there are existing works that proposed measures to analyze user relationships \cite{shahriari2014,gayo2013,jin2013} and behavior (e.g. \textit{retweet}, \textit{mention}, etc.) \cite{khrabrov2010, jabeur2012,aral2012,romero2011,goyal2010,anger2011,cha2010,li2013,silva2013,yamaguchi2010} to find influential users in the OSNs. Yamaguchi et al. \cite{yamaguchi2010} proposed TUrank, which finds influential users in Twitter based on users' links and tweets activities. Silva et al. \cite{silva2013} proposed ProfileRank, which is a PageRank \cite{page1999} inspired model, to find and recommend influential users based on Twitter users' retweet activities. There are also works which extended HITS algorithm \cite{kleinberg1999} to find influential users in OSNs. Romero et al. \cite{romero2011} proposed the influence-passivity (I-P) algorithm to measure Twitter users' influence and passivity from their retweet activities. Gayo-Avello \cite{gayo2013} applied HITS on Twitter follow links to identify and differentiate influential users from spammers. Unlike these works, our paper extends HITS to identify topic-specific hub and authority users in OSNs.  

There are also works that identify topic-specific influential users in OSNs. Many of these works however model the topics and user influence in separate steps  \cite{katsimpras2015,bakshy2011,aleahmad2016,tang2009,weng2010,pal2011,lee2015,hoang2016,hu2013,liu2014,montangero2015}. There are a relatively few works that jointly model user topical interests and influence altogether. Liu et al. \cite{liu2010} proposed a generative model that utilized heterogeneous link information and content to learn the topic-level influence among users. Bi et al. \cite{bi2014} introduced Followship-LDA, a Bernoulli-Multinomial mixture model which jointly learns Twitter users' relationship and tweet content, to identify topic-specific key influential users. Closer to our proposed model, Barbieri et al. \cite{barbieri2014} proposed the WTFW model which models authoritative and susceptible users for different topics. WTFW considers a topic-specific susceptible user as one who is interested in the topic (e.g., posting topic-related content), while a topic-specific authoritative user as one who is followed by many such susceptible users. Differing from WTFW, our proposed model considers the topic-specific hub and authority users, where a topic-specific hub user is one who is not only interested a topic, but also follows many users who are authority in that topic. Conversely, we consider a user an authority for a topic when she is followed by many users who are hubs for the topic. Furthermore, WTFW models authority and susceptibility as distributions of users for different topics, while our proposed model learns the explicit topic-specific hub and authority scores for each user. The explicit scores allow the flexibility for analysis of hubs and authorities across users and topics.

\section{Proposed Model}
\label{(sec:model)}
In this section, we describe our proposed Hub and Authority Topic (HAT) model in detail. We begin by introducing the key elements of the model and their notations. Next, we present the principles behind designing the model and its generative process. We then present an algorithm for learning the model’s parameters and the algorithm's parallelization for speeding up the computation. Lastly, we present a data sub-sampling strategy to further reduce the computational cost.

\subsection{Notations and Preliminaries}

We summarize the main notations in Table~\ref{tab:notations}. We use $U$ and $V$ to denote the sets of followers and followees respectively. For each user $v$, we denote the set of her posts by $S_v$. Here, we adopt the bag-of-word representation for each posts: that is, each post is represented as a multi-set of words, and the word ordering is not important. The number of words of the $s$-th post of user $v$ is then denoted by $N_{v,s}$, while the $n$-th word of the $s$-th post is denoted by $w_{v,s,n}$. Lastly, we denote the number of unique words from all the posts by $W$.

\begin{table}[!t]
	\caption{Notations}
	\label{tab:notations}
	\small
		\begin{tabular}{|c|p{6cm}|}
			\hline
			{\bf Symbol} & {\bf Description}\\			
			\hline
			$U$, $V$& Number of followers/ followees\\
			\hline
			$S_{v}$ & Number of posts of user $v$\\
			\hline
			$N_{v,s}$ & Number of words in post $s_{v}$\\
			\hline
			$W$& Number of unique words\\
			\hline
			$K$& Number of topics\\
			\hline
			$\tau_{k}$ & Word distribution of topic $k$\\
			\hline
			$\theta_{u}$ & Topic distribution of user $u$\\
			\hline
			$A_{v}$ & Topic-specific authority vector of user $u$\\
			\hline
			$H_{u}$ & Topic-specific hub vector of user $u$\\
			\hline
			$r_{uv}$ & Following relationship between $u$ and $v$: $r_{uv} = 1$ if $u$ follows $v$, = 0 otherwise\\
			\hline
			$\alpha$, $\gamma$& Dirichlet priors of $\theta_u$ \& $\tau_k$ respectively\\
			\hline
			$\sigma$, $\delta$& Deviations of $A_v$ \& $H_u$ respectively\\
			\hline
		\end{tabular}
\end{table}

In this work, we adopt topic modeling approach for both users' topical interests and their hubs and authorities specific to each topic. Our proposed model, HAT, consists of the following model elements.

{\bf Topic}. A topic is a semantically coherent theme of words used by users to write posts. Formally, a topic is represented by a multinomial distribution over $W$ (unique) words.

For example, a topic about traveling would have high probabilities for words such as \emph{trip}, \emph{vacation}, and \emph{flight}, but low probabilities for other words. Another topic about food would have high probabilities for words such as \emph{coffee} and \emph{sandwich} but low properties for other non food related words.

{\bf Users' topic distribution}. The topic distribution of a user represents her topical interests or, her preference for different topics. Formally, the topic distribution of user $u$ is a multinomial distribution $\theta_u$ over the set of all topics. We use $k$ to denote the number of topics.

For example, a user interested in traveling would have high probability for traveling topics but low properties for other topics. Similarly, another user interested in fashion would have high probability for fashion related topics but low probabilities for others.

{\bf Topic-specific authority}: This refers to the authority of a user for a specific topic. We want to assign to every user $v$ who is a followee, a topic-specific authority vector $A_v = (A_{v,1}, \cdots,A_{v,K})$ where $K$ is the number of topics and $A_{v,k} \in (0, +\infty)$ for $k=1,\cdots, K$.

{\bf Topic-specific hub}: This refers to the likelihood of a user to be a hub with connection to many authority users for a specific topic. We want to assign to every user $u$ who is a follower, a topic-specific hub vector $H_u = (H_{u,1},\cdots,H_{u,K})$ where $K$ is again the number of topics and $H_{u,k} \in (0, +\infty)$ for $k=1,\cdots, K$.

\subsection{Model Design Principles.}
Our {\bf HAT} model is designed to simulate the process of generating user posts and following links based on their topical interests, hubs, and authorities. We employ topic modeling approach similar to LDA \cite{blei2003} and Twitter-LDA \cite{zhao2011} for generating posts from topics. We also use a factorization approach to generate the following links from topic-specific hubs and authorities. The notable point in our model is in the explicit and direct modeling of the relationships among topical interests, hubs and authorities. We postulate that users' topical interests not only determine post content but also play important roles determining hubs and authorities. The relationship is however not deterministic, but probabilistic in nature. Moreover, users may not be hubs or authorities even in topics they are interested in. Note that our model also learns the explicit scores for users' topic-specific hubs and authorities, which provides the flexibility in analyzing the hubs and authorities across users and topics.

\subsection{Generative Process}

\begin{figure}[!t]
	\begin{center}
		\includegraphics[scale = 0.28]{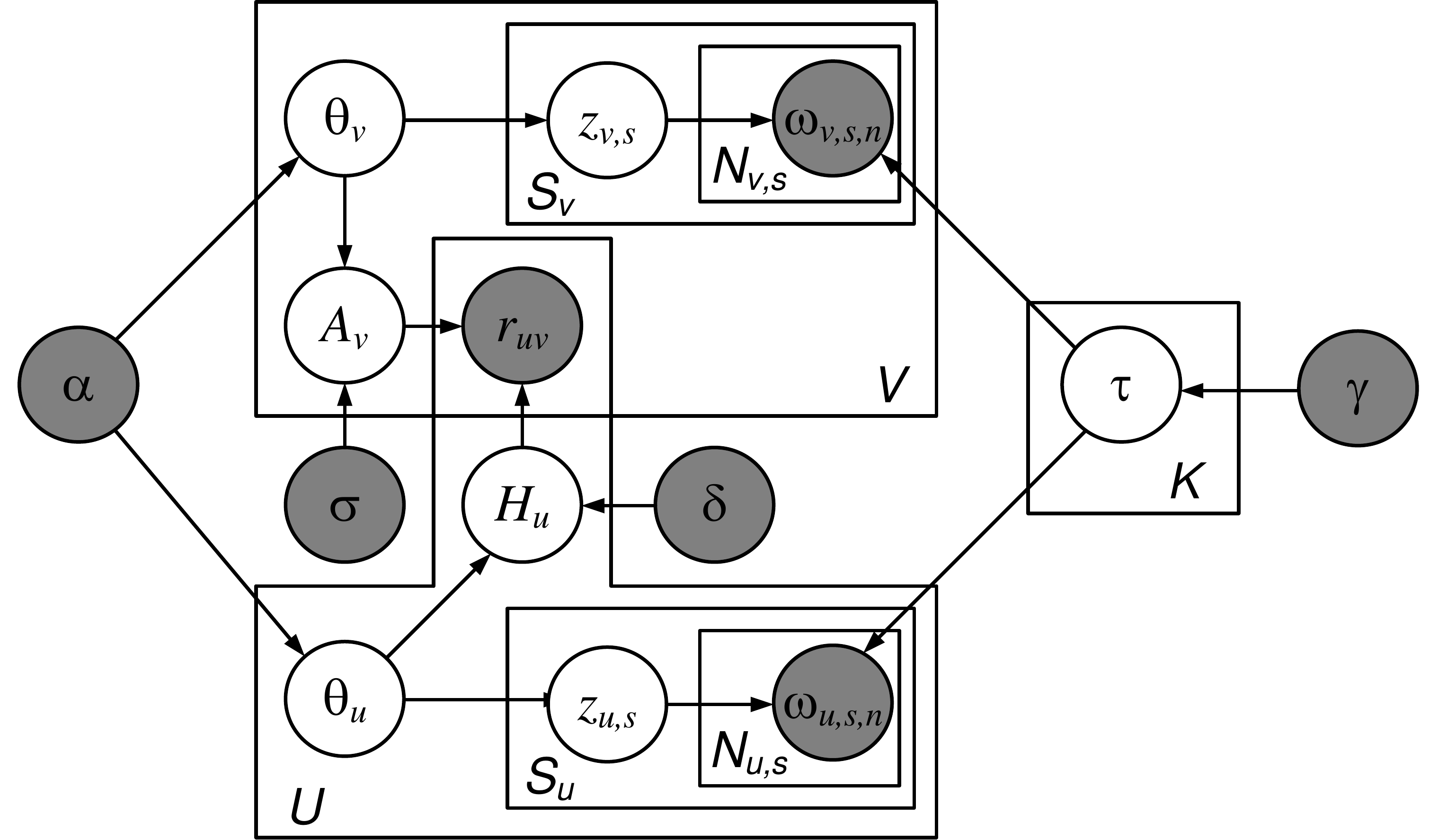}
		\caption{Plate Diagram of HAT Model}
		\label{fig:plate}
	\end{center}
\end{figure}

We depict the plate diagram of the HAT model in Figure \ref{fig:plate}, and summarize its generative process in Algorithm~\ref{algo:Generative}. We first assume that there are $K$ different topics, where $K$ is a given parameters. Each topic $k$ is then a $W$-dimension multinomial distribution $\tau_k$ over $W$ unique words, and is assumed to be sampled from a given Dirichlet prior $\gamma$. For each user $v$, her topic distribution $\theta_v$ is then a $K$-dimension multinomial distribution (over $K$ topics). Similarly, we assumed that $\theta_v$ is sampled from a given Dirichlet prior $\alpha$. User's posts, topic-specific hubs and/or authorities, and following links are then generated as follows.

{\bf Generating posts.} To generate the $s$-th post of user $v$, the post's topic $z_{v,s}$ is first chosen by sampling from $v$'s topic distribution $\theta_{v}$. Similar to previous works on modeling user content in social network \cite{zhao2011}, we assume that each post has only one topic as they are short with limited number of words or characters. The post's content is then generated by sampling its words. Each word $w_{v,s,n}$ is sampled from the topic's word distribution of the chosen topic (i.e., $\tau_{z_{v,s}}$), independently from other words.

{\bf Generating topic-specific hub and authority vectors.} As we want to model the actual values of users' topic-specific hubs and authorities, which are positive numeric values. We also want to probabilistically relate these values to users' topical interests, which are constrained by multinomial distributions. Hence, we propose to use exponential regression approach to model the relationship between topical interests, hubs and authorities. Specifically, the topic-specific authority of user $v$ for topic $k$, $A_{v,k}$, is set to $\exp(x)$, where $x$ is randomly sampled from the normal distribution with mean $\theta_{v,k}$ and deviation $\sigma$. That is, $x \sim N(\theta_{v,k},\sigma)$.  Similarly, the topic-specific hub of user $u$ for topic $k$, $H_{v,k}$, is set to $\exp(y)$, where $y$ is randomly sampled from the normal distribution with mean $\theta_{u,k}$ and deviation $\delta$.  That is, $x \sim N(\theta_{u,k},\delta)$.

{\bf Generating links.} The link is sampled from the Bernoulli distribution with mean $f(H_u^T A_v, \lambda)$ where the function $f(x,\lambda)$ is defined as:

\begin{equation}
\label{dotnorm}
f(x,\lambda) = 2(\frac{1}{e^{-\lambda x} +1} - \frac{1}{2})
\end{equation}
and $\lambda \in (0, 1)$ is an input parameter to scale down $x$.

Next, the likelihood of forming a following link from $u$ to $v$ is factorized into $u$'s topic-specific hubs and $v$'s topic-specific authorities. The likelihood is high when these topic-specific hubs and authorities correlate (i.e., $u$ has high hub in topics that $v$ has high authority), and is low otherwise (e.g., $u$ has high hub values in topics that $v$ has low authority).

\begin{algorithm} [!t]
	\caption{Generative Process for HAT Model}
	\label{algo:Generative}
	\begin{algorithmic} [1]
		\STATE $\boxdot$ ``Generating topics''
		\FOR {each topic $k$}
		\STATE sample the topic's word distribution $\tau_{k} \sim Dir(\gamma)$
		\ENDFOR
		\STATE $\boxdot$ ``Generating users' topical interest''
		\FOR {each user $u$}
		\STATE sample the user topic distribution $\theta_{u} \sim Dir(\alpha)$
		\ENDFOR
		\STATE $\boxdot$ ``Generating users' topic-specific authorities and hubs''
		\FOR {each user $u$}
		\FOR {each topic $k$}
		\STATE sample $x \sim \mathcal{N}(\theta_{u,k},\delta)$
		\STATE $H_{u,k} \gets \exp{(x)}$
		\STATE sample $x \sim \mathcal{N}(\theta_{u,k},\sigma)$
		\STATE $A_{u,k} \gets \exp{(x)}$
		\ENDFOR
		\ENDFOR	
		\STATE $\boxdot$ ``Generating posts''
		\FOR {each user $u$}
		\FOR {each post $s$}
		\STATE sample post's topic $z_{u,s} \sim Multi(\theta_u)$		
		\FOR {each word slot $n$}
		\STATE sample the word $w_{v,s,n} \sim Multi(\tau_{z_{v,s}})$		
		\ENDFOR	
		\ENDFOR	
		\ENDFOR
		\STATE $\boxdot$ ``Generating following relationship''
		\FOR {each pair of follower $u$ and followee $v$}
		\STATE sample the relationship $r_{u,v} \sim Bernoulli(1 - f(H_u^T A_v,\lambda))$ where $f(x,\lambda) = 2(\frac{1}{e^{-\lambda x} +1} - \frac{1}{2})$
		\ENDFOR
	\end{algorithmic}
\end{algorithm}

\subsection{Model learning}
Given the priors $\alpha$ and $\gamma$, and the parameters $\sigma$, $\delta$, and $\lambda$, we learn other parameters in HAT model using maximum likelihood approach. In other words, we solve the following optimization problem.
\begin{multline}
\label{optimization}
\{\theta^{*},A^{*},H^{*},Z^{*},\tau^{*}\} = \\ \text{arg.max}_{\theta,A,H,Z,\tau}L(\text{Data}|\theta,A,H,Z,\tau,\alpha,\gamma,\sigma,\delta,\lambda)
\end{multline}

In Equation~\ref{optimization}, $\theta$ represents for the set of $\theta_u$ for all users $u$. $A$ and $H$ are similarly defined. $Z$ represents for the bag of topics of all posts, while $\tau$ represents for the set of all topic word distributions $\tau_k$'s. Lastly, $L(\text{Data}|\theta,A,H,Z,\tau,\alpha,\gamma,\sigma,\delta,\lambda)$ is the likelihood function of the observed data (i.e., posts and following links) given the value of all the parameters.

Similar to LDA-based models, the problem in Equation~\ref{optimization} is however intractable \cite{blei2003}. We therefore make use of Gibbs-EM method \cite{bilmes1998} for learning in HAT model. More exactly, we first randomly initialize $\theta$, $A$, $H$, and $\tau$. We then iteratively perform the following steps until reaching a convergence or exceeding a given number of iterations.
\begin{itemize}
	\item Gibbs part - to sample $Z$  while fixing $\theta$, $A$, $H$, and $\tau$. The topic $z_{v,s}$ is sampled according to the following equation.
		\begin{equation}
		\label{gibbs}
			P(z_{v,s} = k|\theta_v,\tau) \propto \theta_{v,k}\times \prod_{n=1}^{N_{v,s}}\tau_{k,w_{v,s,n}}
		\end{equation}
	\item EM part - to optimize $\theta$, $A$, $H$, and $\tau$ while keeping $Z$ unchanged. In this step, we make use of the alternating gradient descent method \cite{boyd2017}. That is, we iteratively optimize $\theta$, $A$, $H$, or $\tau$ while fixing all other parameters (and $Z$ as well).
\end{itemize}

\subsection{Parallelization}
As suggested by Equation~\ref{gibbs}, the sampling of a post's topic is independent from that of all the other posts. Hence, we can use multiple child processes, each corresponding to a small set of users, to sample the topics for the users' posts simultaneously. Also, in the alternating steps for optimizing $\theta$, we can parallelize the computation as the optimization of a user's topic distribution is independent of that of all other users' topic distributions. Similarly, we can parallelize the alternating optimization of $A$, $H$, and $\tau$.

In our implementation, in Gibbs-part steps, we build a process pool, and submit a process for sampling topic for posts of $\frac{1}{P}$ of the users where $P$ is the pool's size. In the ideal case, we can reduce the running time of the Gibbs-part to $P$ times. Similarly, we use process pool to reduce the running time in the EM-part's alternating optimization steps.

\subsection{Data sub-sampling}
Like previous factorization and mixed membership models, the HAT model considers both link and non-link relationships of all pair of users. This makes the overall complexity of the HAT model $O(N^2)$ where $N$ is the number of users, which is not practical for large scale social networks. We therefore choose to use a data sub-sampling method for reducing the computational cost. To do that, for each user $u$, we keep all $u$'s out links (i.e., the links where $u$ follows other users) and $p$\% of its out non-links (i.e., the no-links where $u$ does not follows some other users). These $p$\% non-links are selected from the followees of $u$' followees (i.e., the 2-hops non-existing links). These selections allow us to retain only a subset of relationships that carry strong signal of users' hubs and authorities, while filtering out the remaining data that may contain noise. 

\section{Experimental Evaluation}
\label{(sec:experiment}
In this section, we perform experiments to evaluate HAT against state-of-the-art methods. We first introduce two real-world datasets which we have collected for our model evaluation. Next, we describe the experiments conducted and report the results. Finally, we present several empirical findings on the topics, hub and authority users learned by HAT.

\subsection{Datasets}
Our model evaluation requires a dataset that allows us to observe user topical interests and preferences in connecting to authoritative information sources. The requirement is satisfied by two popular social networking platforms, namely Twitter, a short-text microblogging site, and Instagram, a photo-sharing social media site. Both Twitter and Instagram are directed networks, which reflect the preferences of user towards \textit{following} other authoritative users. Furthermore, the hubs and authorities of users in the two platforms may change with respect to different topics.

\begin{table}[H]
	\centering
	\caption{Datasets Statistics}
	\label{tbl:datasets}
	\begin{tabular}{lcc}
		& Instagram                                                                                                             & Twitter                                                                                                                   \\ \cline{2-3} 
		\multicolumn{1}{l|}{\begin{tabular}[c]{@{}l@{}}Total users\\   Total links\\ Avg Links\\  Max followers\\   Max followings\\   Min followers\\   Min following\end{tabular}} & \multicolumn{1}{c|}{\begin{tabular}[c]{@{}c@{}}943\\   33,862\\  35\\   258\\   353\\   4\\   3\end{tabular}} & \multicolumn{1}{c|}{\begin{tabular}[c]{@{}c@{}}9,289\\   316,445\\  35\\   2476\\   899\\   5\\   4\end{tabular}} \\ \cline{2-3} 
		\multicolumn{1}{l|}{\begin{tabular}[c]{@{}l@{}}Total posts\\   Max posts\\   Min posts\\   Avg posts\end{tabular}}                                                                                    & \multicolumn{1}{c|}{\begin{tabular}[c]{@{}c@{}}38,088\\   904\\   5\\   40\end{tabular}}                              & \multicolumn{1}{c|}{\begin{tabular}[c]{@{}c@{}}1,130,632\\   201\\   40\\   121\end{tabular}}                             \\ \cline{2-3} 
	\end{tabular}
\end{table}

For Twitter data, we collected a set of Singapore-based Twitter users who declared Singapore location in their user profiles. These users were identify by an iterative snowball sampling process starting from a small seed set of well known Singapore Twitter users followed by traversing the follow links to other Singapore Twitter users until the sampling iteration did not get any more new users. From these users, we obtain a subset of users who are active, i.e., posted at least 40 tweets, in December 2016. Subsequently, we retrieve the posts of these \textit{active} users published in December 2016. Similar approach is used to retrieve the data of active Instagram users who posted at least 5 photos with captions in December 2016. Table \ref{tbl:datasets} shows the statistics about the collected datasets. In total, we gathered 943 Instagram users and 9,289 Twitter users. There are significantly more Twitter users gathered as many of the Instagram users have private profiles, where their posts are not available. 

\begin{figure*}[t]
	\centering
	\includegraphics[scale = 0.37]{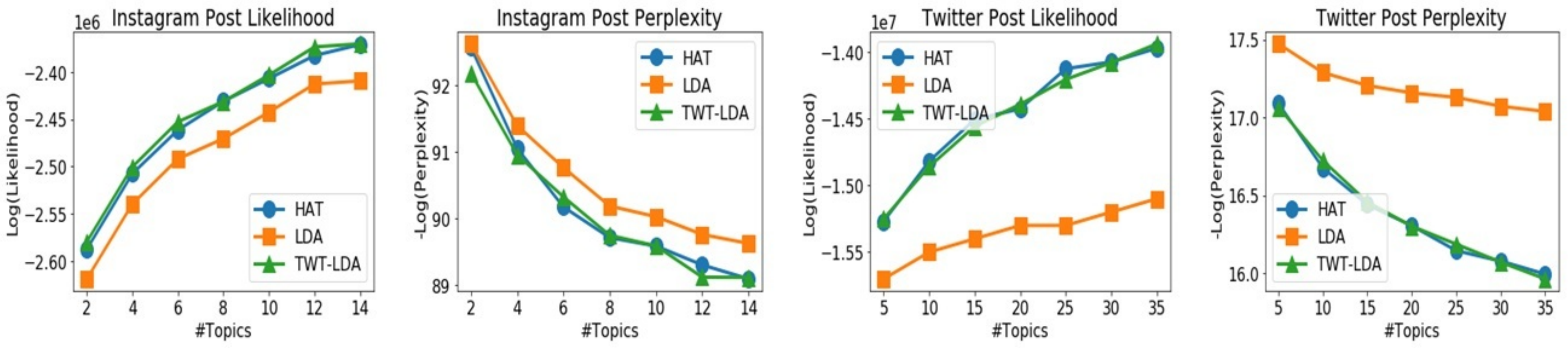}
	\caption{Post Log(Likelihood) and -Log(Perplexity) of HAT, LDA and Twitter-LDA (TW-LDA) for Instagram and Twitter}
	\label{fig:likeplex}
\end{figure*}

\subsection{Experiment Setup}
We evaluate HAT model in two aspects, namely, (i) modeling of topics, and (ii) link recommendation. The former focuses on comparing the topics learned by HAT with those learned by other baseline models.  The latter applies HAT to the prediction of missing links in Twitter and Instagram networks.

\subsubsection{Baselines} For evaluation on topic modeling, we compare HAT with LDA \cite{blei2003} and Twitter-LDA \cite{zhao2011}. The two models are are two popular topic models for text documents and Twitter content respectively.  For link recommendation, we benchmark HAT against several baselines: (i) HITS, (ii) WTFW, and (iii) common user interests learned by LDA and Twitter-LDA. The intuition for interest-based baselines is that user who share common interests are likely to follow each other due to homophily \cite{mcpherson2001}. 

\subsubsection{Training and Test Datasets} In all experiments, we randomly selects 50\% of the posts and links from each user to form the training set, and use the remaining posts and links as the test set. We then learn the HAT and baseline models using the training set, and apply the learned models on the test set. For the link recommendation experiments, we would consider all the links in test set as positive instances, and in principle, all the non-existing links as negative instances. Nevertheless, due to the sparsity of OSNs, the number of possible non-links are enormous. Thus, we limit the negative instances to all the nodes which are 2-hops away from the source node of each positive link.

\subsection{Evaluation on Topic modeling}
In this experiment, we compute the likelihood of the training set and perplexity of the test set when each topic model is applied to the Twitter and Instagram datasets. The model with higher likelihood and lower perplexity is considered superior in this task.

Figure \ref{fig:likeplex} shows the likelihood and perplexity achieved by HAT, LDA and Twitter-LDA. As expected, the larger the number of topics, the higher likelihood and lower perplexity are archived by all models. The quantum of improvement, however, reduces as the number of topics increases. 

Figure \ref{fig:likeplex} also shows that HAT outperforms LDA, and is comparable to Twitter-LDA in the topic modeling task. This result supports the insights from previous work which suggested that standard LDA does not work well for short social media text as both Instagram photo captions and Twitter tweets are much shorter than normal documents \cite{zhao2011}. A possible explanation for the similar results achieved by HAT and Twitter-LDA can be due to both models assuming that each post has only one topic.

Interestingly, we also observe that HAT and Twitter-LDA have outperformed LDA more in Twitter than Instagram. A possible explanation can again be attributed the different length of the post in different OSNs; Twitter tweets as shorter with a 140 character limit, while Instagram photo captions are longer with no limitation in length imposed. 

\subsection{Evaluation on Link Recommendation}
We define the link recommendation task as recommending new links to user, i.e., we want to recommend users other users to follow. Thus, given a user $u$, we first rank her positive and negative instances in test set by some link scores. Then, we recommend $u$ other users $v$ who are higher on the link scores.

\begin{table}[!t]
	\centering
	\footnotesize
	\caption{Instagram link recommendation results}
	\label{tbl:instagram_results}
	\begin{tabular}{lccccc}
		& \multicolumn{4}{c}{Precision @ Top k}                                                                                                                 & \multicolumn{1}{l}{}                \\ \cline{2-6} 
		\multicolumn{1}{c|}{Methods}  & \multicolumn{1}{c|}{k=1}              & \multicolumn{1}{c|}{k=2}              & \multicolumn{1}{c|}{k=3}              & \multicolumn{1}{c|}{k=4}              & \multicolumn{1}{c|}{MRR}            \\ \hline
		\multicolumn{1}{|l|}{LDA}     & \multicolumn{1}{c|}{0.088}          & \multicolumn{1}{c|}{0.101}          & \multicolumn{1}{c|}{0.107}          & \multicolumn{1}{c|}{0.115}          & \multicolumn{1}{c|}{0.223}          \\ \hline
		\multicolumn{1}{|l|}{TWT-LDA} & \multicolumn{1}{c|}{0.143}          & \multicolumn{1}{c|}{0.145}          & \multicolumn{1}{c|}{0.143}          & \multicolumn{1}{c|}{0.145}          & \multicolumn{1}{c|}{0.279}          \\ \hline
		\multicolumn{1}{|l|}{HITS}    & \multicolumn{1}{c|}{0.234}          & \multicolumn{1}{c|}{0.203}          & \multicolumn{1}{c|}{0.210}          & \multicolumn{1}{c|}{0.220}          & \multicolumn{1}{c|}{0.379}          \\ \hline
		\multicolumn{1}{|l|}{WTFW}    & \multicolumn{1}{c|}{0.339}          & \multicolumn{1}{c|}{0.321}          & \multicolumn{1}{c|}{0.310}          & \multicolumn{1}{c|}{0.313}          & \multicolumn{1}{c|}{0.496}          \\ \hline
		\multicolumn{1}{|l|}{HAT}     & \multicolumn{1}{c|}{\textbf{0.462}} & \multicolumn{1}{c|}{\textbf{0.408}} & \multicolumn{1}{c|}{\textbf{0.400}} & \multicolumn{1}{c|}{\textbf{0.396}} & \multicolumn{1}{c|}{\textbf{0.597}} \\ \hline
	\end{tabular}
\end{table}
%\vspace{-10px}
\begin{table}[!t]
	\centering
	\footnotesize
	\caption{Twitter link recommendation results}
	\label{tbl:twitter_results}
	\begin{tabular}{lccccc}
		& \multicolumn{4}{c}{Precision @ Top k}                                                                                                                 & \multicolumn{1}{l}{}                \\ \cline{2-6} 
		\multicolumn{1}{c|}{Methods}  & \multicolumn{1}{c|}{k=1}              & \multicolumn{1}{c|}{k=2}              & \multicolumn{1}{c|}{k=3}              & \multicolumn{1}{c|}{k=4}              & \multicolumn{1}{c|}{MRR}            \\ \hline
		\multicolumn{1}{|l|}{LDA}     & \multicolumn{1}{c|}{0.126}          & \multicolumn{1}{c|}{0.125}          & \multicolumn{1}{c|}{0.120}          & \multicolumn{1}{c|}{0.118}          & \multicolumn{1}{c|}{0.261}          \\ \hline
		\multicolumn{1}{|l|}{TWT-LDA} & \multicolumn{1}{c|}{0.144}          & \multicolumn{1}{c|}{0.141}          & \multicolumn{1}{c|}{0.136}          & \multicolumn{1}{c|}{0.131}          & \multicolumn{1}{c|}{0.279}          \\ \hline
		\multicolumn{1}{|l|}{HITS}    & \multicolumn{1}{c|}{0.247}          & \multicolumn{1}{c|}{0.236}          & \multicolumn{1}{c|}{0.229}          & \multicolumn{1}{c|}{0.221}          & \multicolumn{1}{c|}{0.408}          \\ \hline
		\multicolumn{1}{|l|}{WTFW}    & \multicolumn{1}{c|}{0.288}          & \multicolumn{1}{c|}{0.261}          & \multicolumn{1}{c|}{0.239}          & \multicolumn{1}{c|}{0.227}          & \multicolumn{1}{c|}{0.452}          \\ \hline
		\multicolumn{1}{|l|}{HAT}     & \multicolumn{1}{c|}{\textbf{0.452}} & \multicolumn{1}{c|}{\textbf{0.397}} & \multicolumn{1}{c|}{\textbf{0.358}} & \multicolumn{1}{c|}{\textbf{0.330}} & \multicolumn{1}{c|}{\textbf{0.572}} \\ \hline
	\end{tabular}
\end{table}

To compute a link score, $score(u,v)$, between two users using HAT, we take the inner product of source user $u$'s topic-specific hub scores, $H_u$, and target user $v$'s authority scores topic-specific authority scores, $A_v$:
\
\begin{equation}
\label{eqn:hat_predict}
score(u,v) = \sum^{K}_{k=1} H_{u,k} \cdot  A_{v,k}
\end{equation}

For HITS, the link score for a link is compute by taking the product of source user's hub ($h_u$) and target user's authority ($a_v$):

\begin{equation}
\label{eqn:hits_predict}
score(u,v) = h_{u} \cdot  a_{v}
\end{equation}
We also compute the link score for a link using the user interests learned from LDA and Twitter-LDA, i.e., we compute the link score of a source user $u$ linking to target user $v$ by taking the inner product of the topical interests, $\theta$, of $u$ and $v$:

\begin{equation}
\label{eqn:interest_predict}
score(u,v) = \sum^{K}_{k=1} \theta_{u,k} \cdot  \theta_{v,k}
\end{equation}

The WTFW model is designed to returns a link score for a given link based on the parameters learned by the model. 

Finally, we use \textit{precision at top k} and Mean Reciprocal Rank (MRR) \cite{voorhees2003} to measure the accuracy of link recommendation. \textit{Precision at top k} is defined as:

\begin{multline*}
Prec_k = \frac{\parbox[c]{12em}{\#users with positive links recommended in top k}}{\text{\#users with at least k positive links}}
\end{multline*}

Tables \ref{tbl:instagram_results} and \ref{tbl:twitter_results} show the link recommendation results for Instagram and Twitter respectively. Note that for HAT and the topic-specific baselines, i.e., WTFW, LDA and Twitter-LDA, the number of topic learned in training phase is set to 4 and 15 for Instagram and Twitter respectively. 

We observe that HAT outperforms all baselines measured by \textit{both precision at top k} and MRR for both Instagram and Twitter. When measured by MRR, HAT has significantly outperformed HITS by more than 50\% and 40\% in Instagram and Twitter respectively. This suggests that the topical context is important in link recommendation. HAT also improves the MRR of the common user interests baselines by more than one fold. This also suggests the importance of network information in recommending users to follow. 

Finally, considering both OSNs, HAT also outperforms WTFW by 20\% in MRR. Interestingly, this demonstrates the importance of hub when modeling topical links; WTFW models susceptibility as users who are interested in a particular topic, while HAT models hub as users who are not only interested in a topic but follow users who are also authority users in that topic.

\begin{table*} [t]
	\caption{A sample of topical authority and hub users in Instagram and Twitter learned by HAT}
	\label{tab:hat_emp}
	\footnotesize
	\centering
	\begin{tabular}{|p{1.5cm}|p{4cm}|p{5cm}|p{5cm}|}
		\hline
		Topic & Top 10 Keywords & Top 5 Authority Users & Top 5 Hub Users \\ \hline \hline
		\multicolumn{4}{|c|}{\textbf{Instagram}} \\ \hline
		Fitness & fitness, build, training, coach, gymnasticbodies, gym, gymnastics, bones, session, muscles & \texttt{danielchan\char`_gymnasticbodies, iron\char`_fitness\char`_Singapore (gym), stepdancesg (dance studio), heechai (gym enthusiast), cheryltaysg (fitness blogger)} & \texttt{iron\char`_fitness\char`_Singapore (gym), cheryltaysg (fitness blogger), \char`_yaops\char`_(athlete),  herworldsingapore (lifestyle magazine), rei\char`_angeline (model, fitness enthusiast)} \\ \hline
		Fashion & runwayspree, stylexstyle, ootd, wiwt, fashion, spree, clozette, dress, outfit, style & \texttt{clozette, thestagewalk (online boutiques), crystalphuong, linyuhsinx (lifestyle bloggers), aanurul (actress)} & \texttt{thequeenapple, linyuhsinx, crystalphuong, francescasoh (lifestyle bloggers), madameflamingo (online boutiques)} \\ \hline
		Gourmet & food, foodporn, foodstagram, delicious, sgfood, instafood, foodgasm, foodpic, eatoutsg, sgfoodies & \texttt{misstamchiak, ieatishootipost, danielfooddiary, hungrygowhere, sgfoodonfoot (food bloggers)} & \texttt{eatdreamlove (food blogger), cahskb, bobcatsysop (average users, food lovers), tessbarsg, drcafesg (cafes)} \\ \hline \hline
		\multicolumn{4}{|c|}{\textbf{Twitter}} \\ \hline
		Current Affairs & singapore, trump, stcom, Asian, south, china, channelnewasia, world, centre & \texttt{stcom, channelnewsasia, stompsingapore (Singapore news media), leehsienloong (Singapore Prime Minister), sgag\char`_sg (satire media)} & \texttt{herbertrsim (columnist), mrbrown (satirical blogger) miyagi (comedian) , lkysch (public policy school), anitakapoor(TV host)} \\ \hline
		\texttt{K-Pop} & bigolive, bts, btstwt, allkpop, exo, music, baekhyun, stage, tour, bigbang & \texttt{hallyusg, sgxclusive, kavenyou (K-pop news media), exonationsg, sgvips (K-pop band fan group)} & \texttt{jessicaxtanx, michellensm, nuraeffy, adelinee\char`_yap, happypeachniel (average user, k-pop fans)} \\ \hline
		Football & manutd, lfc, geniusfootball, mufc, arsenal, football, goals, league, team, mourinho & \texttt{elevensportssg, fourfourtwosg (sport media), serisyarianie, aliedrvs (football enthusiasts), mightystags (Singapore football club)} & \texttt{zionsport (football field rental), elevensportssg (sport media), heyiamqayyum, serisyarianie, aliedrvs (football enthusiasts)} \\ \hline
	\end{tabular}
	\normalsize
\end{table*}

\begin{table} [h]
	\caption{A sample of authority and hub users in Instagram and Twitter learned by HITS}
	\label{tab:hits_emp}
	\centering
	\footnotesize
	\begin{tabular}{|p{3.6cm}|p{3.6cm}|}
		\hline
		Top 5 Authority Users & Top 5 Hub Users \\ \hline \hline
		\multicolumn{2}{|c|}{\textbf{Instagram}} \\ \hline
		misstamchiak, hungrygowhere (food blogger),  superadrianme (lifestyle blogger), geraldpng (model), thesingaporewomensweekly (lifestyle magazine) & misstamchiak (food blogger), a.mandayong, superadrianme (lifestyle blogger), geraldpng (model) theinfluencernetwork (media company) \\ \hline \hline
		\multicolumn{2}{|c|}{\textbf{Twitter}} \\ \hline 
		T\texttt{xavierlur, syakirahnasri, naomineo\char`_ (lifestyle blogger), sgag\char`_sg (satire media), mrbrown (satirical blogger)} & \texttt{xavierlur, syakirahnasri, \char`_nadd24, yourh1ghne5s (lifestyle blogger), mrbrown (satirical blogger)} \\ \hline
	\end{tabular}
	\normalsize
\end{table}

\subsection{Empirical Analysis}
In this section, we empirically compare the authority and hub users learned by HAT and HITS. Tables \ref{tab:hat_emp} and \ref{tab:hits_emp} show samples of the authority and hub users in Instagram and Twitter learned by HAT and HITS respectively. HITS basically determines the authority and hub users strictly by the network structures. Thus, the top authority and hub users identified by HITS are popular Twitter and Instagram users with many followers. On the other hand, HAT is able to identify authority and hub users for specific topic. For example, for the topic on "Fashion`` in Instagram, HAT was able to identify popular online boutiques, celebrity lifestyle bloggers and an actress as top authority users. These users often post fashion-related content and are followed by many users who are interested in fashion. Similarly, the top fashion topic hub users identified by HAT are also lifestyle bloggers and online boutiques who have followed the fashion topic authority users. Similar observations are made for Twitter.    

There are also some insightful findings when analyzing the authority and hub users suggested by HAT. For example, we observe that the top authority users for ``current affair" are mainly the Singapore mainstream media and the current Prime Minister for Singapore. However, the top hub users for this topic are more diverse, which consist of columnist, comedian and satirical blogger. The ``k-pop" topic in Twitter also has interesting authority and hub users; the top authority users are mainly k-pop news media and k-pop band fan groups, while the top hub users are k-pop fans with average number of followers. This suggests that for the k-pop topic, most of the interested users are connected to many authority users, thus making the hub scores for this topic less relevant.    

\section{Conclusion}
\label{(sec:conclusion}
In this paper, we have proposed a novel generative model called Hub and Authority Topic (HAT) model, which jointly models user's topical interests, hubs and authorities. We evaluated HAT using real-world datasets and benchmarked against the state-of-the-art. Our experiments have shown that HAT outperforms LDA and achieve comparable results as Twitter-LDA in topic modeling. On link recommendation, HAT outperforms the baseline methods in MRR by at least 20\%. We have also empirically shown that HAT is able to identify hub and authority users for different topics in Twitter and Instagram. For future works, we would like extend our model to include non-topical relationship among users. Currently, our model assumes that all links among users are topical, however, a user may follow each other for social reasons (e.g., they are friends).

\balance
\bibliographystyle{siam}
\bibliography{ref}  
\end{document}